\documentclass[12pt,a4paper]{iopart} 
\usepackage{iopams}
\usepackage{setstack,cite}
\usepackage{amssymb,graphicx}
  
\begin{document}

\title[]{Multihumped nondegenerate fundamental bright solitons in $N$-coupled nonlinear Schr\"{o}dinger system}
\author{R. Ramakrishnan, S. Stalin$\dagger$, M. Lakshmanan$\ddagger$}
\address{Department of Nonlinear Dynamics, Bharathidasan University, Tiruchirappalli - 620 024, Tamilnadu, India.}
\ead{$\ddagger$ lakshman.cnld@gmail.com}
\ead{$\dagger$stalin.cnld@gmail.com (Corresponding Author)}

\begin{abstract}
In this letter we report the existence of nondegenerate fundamental bright soliton solution for coupled multi-component nonlinear Schr\"{o}dinger equations of Manakov type. To derive this class of nondegenerate vector soliton solutions, we adopt the Hirota bilinear method with appopriate general class of seed solutions. Very interestingly the obtained nondegenerate fundamental soliton solution of the $N$-coupled nonlinear Schr\"{o}dinger (CNLS) system admits multi-hump natured intensity profiles. We explicitly demonstrate this specific property by considering the nondegenerate soliton solutions for $3$ and $4$-CNLS systems. We also point out the existence of a special class of partially nondegenerate soliton solutions by imposing appropriate restrictions on the wavenumbers in the already obtained completely nondegenerate soliton solution. Such  class of soliton solutions can also exhibit multi-hump profile structures. Finally, we present the stability analysis of nondegenerate fundamental soliton of the $3$-CNLS system as an example. The numerical results confirm the stability of triple-humped profile nature  against perturbations of   5\% and 10\% white noise. The multi-hump nature of nondegenerate fundamental soliton solution will be usefull in multi-level optical communication applications with enhanced flow of data in multi-mode fibers.  
\end{abstract}

Multi-level optical communication with high bit-rate data transmission is a hotly debated topic and is a challenging task in optical communication applications. Using wavelength division multiplexing scheme, the conventional binary data transmission approaches its limit \cite{Rohrmann}, where the maximum data-carrying rate of the fiber is restricted by Shannon's theorem \cite{sha} due to channel capacity crunch. In the conventional binary data coding, the presence of light pulse is represented by logical ``1" and logical ``0" corresponds to its absence. However, the demand for fiber's information carrying capacity is increasing day by day. To improve the underlying technology it has been proposed that soliton assisted  fiber-optic telecommunication will play a crucial role in determining the future communication systems. Several coding schemes have been proposed in the past to develop this technology: For examble, solitons \cite{td}, which are stable localized nonlinear wave solutions of nonlinear Schr\"{o}dinger  equation, are being proposed as constituting a model for optical pulse propagation in fibers as natural bits for coding the information. Recently, the existence of soliton molecules in dispersion-managed fiber \cite{fm1} has been demonstrated and their possible usefulness in optical telecommunications technology with enhanced data carrying capacity has been pointed out \cite{fm2}. Soliton molecule is a bound soliton state which can be formed when two antiphase solitons persist at a stable equilibrium separation distance, where the interaction force is null among the individuals. Such stable equilibrium  manifests this bound state structure,  reminiscent of a diatomic molecule in condensed matter physics.   The binding force arises between the constituents of the soliton compound due to the Kerr nonlinearity \cite{gord,fm3} and the detailed mechanism can be found in Ref. \cite{fm4}. The existence of two-pulse and three-pulse molecules complete the next level of alphabet of symbols. Such soliton molecules allow coding of two-bits of information simultanously in a single time slot. In this way, the soliton molecules increase the flow of data in fibers. It should be noted here that the initial shape (symmetric peaks with equal intensities) of solion molecules changes due to various losses in the fiber and its intrinsic nonlinearities. However, their fundamental properties do not change during the evolution. Apart from the above, the concept of soliton molecules has been discussed earlier in detail in the context of non-dispersion managed fiber \cite{sm1,sm2,sm3} and in fiber lasers \cite{fl1,fl2,fl3}. In addition to the above, multi-soliton complexes in multimode fibers have also been discussed for increasing the bit-rate in multi-level coding of information \cite{na1,na2,na3}. 

Very recently we have identified a new class of nondegenerate vector bright solitons \cite{Stalin1}, with double-hump nature characterized by two distinct wavenumbers, for the Manakov system \cite{Manakov}. Basically the Manakov system is a model for propagation of orthogonally polarized optical waves in birefringent fiber, where the solitons undergo collision without energy redistribution in general among the modes depending upon the choice of soliton parameters \cite{Stalin1,Qin}. However, they encounter shape changing collision for suitable choice of parameters whenever they interact with themselves or when they collide with degenerate vector brights solitons, that is solitons with single-peak intensity profile described by identical wavenumbers in both the modes \cite{Stalin2}. Such nondegenerate solitons (NDSs) exhibit multi-hump profiles, as we describe below in the present letter, in the case of $N$-CNLS system which may be relevent for optical communication applications. By exploiting the multi-peaks, with different peak powers, the nature of NDSs can be made useful to code the two bits of information as described in \cite{Rohrmann} in the next level of binary coding. To the best of our knowledge study on  nondegenerate solitons in multi-mode fibers or fiber arrays is missing in the literature and their existence in multi-component nonlinear Schr\"{o}dinger system and their usefulness in the context of higher bit-rate information transmission applications have not been reported.  In addition, the underlying interesting analytical forms of NDSs and their geometrical profiles have not been revealed so far in the literature and they need to be analysed in detail.    

In this letter, we intend to investigate the multi-hump nature of nondegenerate fundamental solitons in the following system of multi-component nonlinear Schr\"{o}dinger equations  
\begin{equation}
\label{eq1}
iq_{j,z}+q_{j,tt}+2 \sum_{p=1}^{N}|q_{p}|^{2}q_{j}=0,~~~j=1,2,...,N,
\end{equation}
 by deriving their analytical forms through Hirota bilinear method. Equation (\ref{eq1}) describes the optical pulse propagation in $N$-mode optical fibers \cite{gpa} and it describes the incoherent light beam propagation in photorefractive medium \cite{na2} and so on. In the above, $q_{j}$'s are complex wave envelopes propagating in $N$-optical modes and $z$ and $t$ represent the normalized distance and retarded time, respectively.  We note that for $N=2$ in Eq. (\ref{eq1}), we have studied the collision and stability properties of the nondegenerate solitons \cite{Stalin2} and also we have identified their existence in other integrable nonlinear Schr\"{o}dinger family of equations by revealing their analytical forms \cite{Stalin3}. To derive the exact form of the nondegenerate fundamental soliton solution for the $N$-CNLS sytem, we bilinearize Eq. (\ref{eq1}) through the dependent variable transformation, $q_{j}(z,t)=\frac{g^{(j)}(z,t)}{f(z,t)}$, $j=1,2,...,N$ where $g^{(j)}$'s are in general complex functions and $f$ is a real function. Substitution of this transformation in Eq. (\ref{eq1}) brings out the following bilinear forms: $(iD_{z}+D_{t}^{2})g^{(j)}\cdot f=0$ and $D_{t}^{2}f \cdot f=2(\sum_{n=1}^{N}g^{(n)}\cdot g^{(n)*})$. Here $D_{z}$ and $D_{t}$ are the usual Hirota bilinear operators \cite{Hirota}. Then we consider the standard Hirota series expansions $g^{(j)}=\epsilon g_1^{(j)}+\epsilon^3 g_3^{(j)}+...$, $j=1,2,...,N$ and  $f
 =1+\epsilon^2 f_2+\epsilon^4 f_4+...$ in the solution construction process. 
 
 To obtain the nondegenerate fundamental soliton solution of Eq. (\ref{eq1}) we consider the general forms of $N$-seed solutions, $g^{(j)}=\alpha_{1}^{(j)}e^{\eta_{j}}$, $\eta_{j}=k_{j}t+ik_{j}^{2}z$, where $\alpha_{1}^{(j)}$ and $k_j$, $j=1,2,...,N$ are complex parameters and are nonidentical in general to the $N$-independent linear partial differential equations, $ig_{1,z}^{(j)}+g_{1,tt}^{(j)}=0$, $j=1,2,...,N$, which arise at the lowest order of $\epsilon$.  With such general choices of seed solutions, we proceed to solve the resulting inhomogeneous linear partial differential equations successively in order to deduce the full series solution upto  $g_{2N-1}^{(j)}$ in $g^{(j)}$ and $f_{2N}$ in $f$. By combining the obtained forms of the unknown functions as per the series expansions we find a rather complicated form of the nondegenerate fundamental soliton solution for the $N$-CNLS equation. However, we have managed to rewrite it in a more compact form using the following Gram determinants \cite{Ablowitz, Vijayajayanthi}, \begin{eqnarray}
 \label{eq2}
 g^{(N)} =
 \left|
 \begin{array}{ccc}
 A & I & \phi \\
 -I & B & \bf{0}^{T} \\
 \bf{0} & C_{N} & 0
\end{array}
\right|,~~~
 f =
  \left|
 \begin{array}{cc}
 A & I  \\
 -I & B   
\end{array}
\right|,
  \end{eqnarray}
 where the elements of the matrices $A$ and $B$ are
 \begin{eqnarray}
 &&\hspace{-1.7cm} A_{ij}=\frac{e^{\eta_{i}+\eta_{j}^{*}}}{(k_{i}+k_{j}^{*})}, ~ B_{ij}=\kappa_{ji}=\frac{\psi_{i}^{\dagger} \sigma \psi_{j}}{(k_{i}^{*}+k_{j})},~
 C_{N} = - 
\left(
\begin{array}{ccccc}
 \alpha_{1}^{(1)}, & \alpha_{1}^{(2)}, & .~.~.~,& \alpha_{1}^{(N)} 
\end{array}
\right),\nonumber \\
 &&\hspace{-1.7cm}\psi_{j}=
\left(
\begin{array}{ccccc}
 \alpha_{1}^{(1)}, &\alpha_{1}^{(2)},& .~ .~ . ~,&\alpha_{1}^{(j)} 
\end{array}
\right)^T,
 \phi=
\left(
\begin{array}{ccccc}
 e^{\eta_{1}},&	e^{\eta_{2}},&. ~.~.~, &e^{\eta_{n}}  
\end{array}
\right)^T, j,n=1,2,..,N.\nonumber
 \end{eqnarray}     
In the above, $g^{(N)}$ and $f$ are $((2^2N)+1)$ and  $(2^2N)$th order determinants, $T$  represents the transpose of the matrices $\psi_{j}$ and $\phi$, $\dagger$ denotes transpose complex conjugate,  $\sigma=I$ is an (n $\times$ n) identity matrix, $\phi$ denotes (n $\times$ 1) column matrix, $\bf{0}$ is a (1 $\times$ n) null matrix, $C_{N}$ is a (1 $\times$ n) row matrix  and  $\psi$ represents a (n $\times$ 1) column matrix. In the above expressions, for the nondegenerate fundamental soliton solution  the elements $\kappa_{ji}$'s do not exist ($\kappa_{ji}=0$) in the square matrix $B$ when $j \neq i$. Also for a given set of $N$ and $j$ values the corresponding elements only exist and all the other elements are equal to zero in $C_{N}$ and  $\psi_{j}$ matrices (we have demonstrated the latter clearly for the $3$-component case below). We have verified the validity of the nondegenerate fundamental soliton solution (\ref{eq2}) by substituting it in the bilinear equations of Eq. (\ref{eq1}) along with the following derivative formula of the determinants, $\frac{\partial M}{\partial x}	= \sum_{1\le i,j\le n} \frac{\partial a_{i,j}}{\partial x} \frac{\partial M }{\partial a_{i,j}} = \sum_{1\le i,j\le n} \frac{\partial a_{i,j}}{\partial x} \Delta_{i,j}$, where $\Delta_{i,j}$'s are the cofactors of the matrix $M$, the bordered determinant properties and the elementary properties of the determinants \cite{Hirota}. This action yields a pair of Jacobi identities and thus their occurrence confirms the validity of the obtained soliton solution. Multi-hump profile nature is a special feature of the obtained nondegenerate fundamental soliton solution (\ref{eq2}). Such multi-hump structures and their propagation are characterized by $2N$ arbitrary complex wave parameters. The funamental nondegenerate soliton admits a very interesting $N$-hump profile in the present $N$-CNLS system. In this system, in general, the nondegenerate solitons propagate with different velocities in different modes but one can make them to propagate with identical velocity by restricting the imaginary parts of all the wave numbers $k_{j}$, $j=1,2,...,N$, to be equal.  Interestingly, in 1976, Nogami and Warke have obtained  soliton solution for the multicomponent CNLS system \cite{nogami}.   We note that their soliton solution corresponds to the so called partially coherent soliton (PCS) which can be checked after replacing the function $e_j=\exp(k_jx)$ by $e_j=\sqrt{2k_ja_j}\exp(k_j\bar{x}_j)$, where $\bar{x}_j=x-x_j$, $a_j=\Pi_{j\neq i} c_{ij}$, $c_{ij}=\frac{k_i+k_j}{|k_i-k_j|}$ and $k_j$'s are real constants, in their solution \cite{akhmediev}. Since, the stationary $N$-PCS solution arises from our solution (2) under the parametric restrictions $\alpha_1^{(j)}=e^{\eta_{j0}}$, $j=1,3,4,...,N$ and $\alpha_1^{(2)}=-e^{\eta_{20}}$, ($\eta_{j0}$: real), $k_j=k_{jR}$, $k_{jI}=0$, $j=1,2,...,N$, the solution of Nogami and Warke \cite{nogami} and its time dependent version are essentially special cases of our general solution (\ref{eq2}).    

It is interesting to note that if we set all the wavenumbers $k_{j}$, $j=1,2,...,N$, as identical, $k_j=k_1$, $j=1,2,...,N$, which corresponds to the seed solutions getting restricted as $g^{(j)}=\alpha_{1}^{(j)}e^{\eta_{1}}$, $\eta_{1}=k_{1}t+ik_{1}^{2}z$, for all $j=1,2,...,N,$  in the fundamental soliton solution (\ref{eq2}), the resultant form gets reduced to the following degenerate soliton (DS) solution for Eq. (\ref{eq1}) \cite{Kanna3} as \begin{eqnarray}
\left(
q_1,
q_2, 
q_3,
...,q_N
\right)^T
& = & \left(
A_1,
A_2,
A_3,
...,A_N
\right)^T
k_{1R}e^{i\eta_{1I}}{\mbox{sech}
	\left(\eta_{1R}+\frac{R}{2}\right)}, \label{deg}
\end{eqnarray}
where $\eta_{1R}=k_{1R}(t-2k_{1I}z)$, $A_j=\alpha_1^{(j)}/
\Delta$
and
$\Delta=
((\sum_{j=1}^N{|\alpha_1^{(j)}|^2}))^{1/2}$. 
Here $\alpha_1^{(j)}$, $k_1$, $j=1,2,...,N$, are arbitrary complex
parameters. Further, $k_{1R}A_j$ gives the amplitude of the $j$th mode, $\frac{R}{2}(=\frac{1}{2}\log\frac{\Delta}{(k_1+k_1^*)^2})$ denotes the central position of the soliton
and $2k_{1I}$ is the soliton velocity \cite{Kanna3}. It is evident that the degenerate soliton solution (\ref{deg}) always admits single-hump structure. Using this single peak intensity or power profile as signal in binary coding one cannot improve higher bit-rate in  information transmission as pointed out in \cite{fm1} whereas this class of degenerate solitons interestingly exhibit energy exhanging collision  leading to the construction of all optical logic gates \cite{ma}. To enhance the bit-rate multi-hump pulses with symmetric and asymmetric profiles, as we describe below for $3$ and $4$-CNLS systems as examples,  can be useful for optical communication. 

In order to show the multi-hump nature of the nondegenerate soliton, here we demonstrate such special feature in the case of $3$-CNLS and $4$-CNLS systems. To start with, we consider the three coupled nonlinear Schr\"{o}dinger equation ($N=3$ in Eq. (\ref{eq1})). To get the nondegenerate fundamental soliton solution for this system, we consider the solutions, $g_{1}^{(1)}=\alpha_{1}^{(1)}e^{\eta_{1}}$, $g_{1}^{(2)}=\alpha_{1}^{(2)}e^{\eta_{2}}$ and $g_{1}^{(3)}=\alpha_{1}^{(3)}e^{\eta_{3}}$ as seed solutions to the lowest order linear PDEs. These general form of seed solutions terminates the series expansions as $g^{(j)}=\epsilon g_1^{(j)}+\epsilon^3 g_3^{(j)}+\epsilon^5 g_5^{(j)}$, $j=1,2,3$ and  $f
=1+\epsilon^2 f_2+\epsilon^4 f_4+\epsilon^6 f_6$. By rewriting the explicit forms of the obtained unknown functions in terms of Gram determinants we get the resultant forms similar to the one (Eq. (\ref{eq2})) reported above for the $N$-component case. We find that for  the $3$-CNLS system the matrices $A$ and $B$ are constituted by the elements, $A_{ij}$ and $B_{ij}$, $i,j=1,2,3$ and also the other matrices $C_N$, $\psi_j$ and $\phi$ are deduced as $C_1=\left(
\begin{array}{ccccc}
\alpha_{1}^{(1)} & 0 & 0
\end{array}
\right)$, $C_2=\left(
\begin{array}{ccccc}
0 & \alpha_{1}^{(2)} & 0
\end{array}
\right)$, $C_3=\left(
\begin{array}{ccccc}
0 & 0 & \alpha_{1}^{(3)}
\end{array}
\right)$, $\psi_{1}=
\left(
\begin{array}{ccccc}
\alpha_{1}^{(1)} &0& 0
\end{array}
\right)^T$, $\psi_{2}=
\left(
\begin{array}{ccccc}
0 &\alpha_{1}^{(2)}& 0
\end{array}
\right)^T$, $\psi_{3}=
\left(
\begin{array}{ccccc}
0 &0& \alpha_{1}^{(3)}
\end{array}
\right)^T$ and $\phi=
\left(
\begin{array}{ccccc}
e^{\eta_{1}}&	e^{\eta_{2}} &e^{\eta_{3}}  
\end{array}
\right)^T$. From the resultant Gram-determinant forms, we deduce the following triple-humped nondegenerate fundamental soliton solution for the $3$-CNLS system,
\begin{eqnarray}
\hspace{-2.5cm}q_1&=&\frac{1}{f}e^{i\eta_{1I}}\bigg(e^{\frac{\Delta_{51}+\rho_{11}}{2}}\cosh(\eta_{2R}+\eta_{3R}+\frac{\phi_1}{2})+e^{\frac{\Delta_{11}+\Delta_{21}}{2}}\cosh(\eta_{2R}-\eta_{3R}+\frac{\phi_2}{2})\bigg),\nonumber\\
\hspace{-2.5cm}q_2&=&\frac{1}{f}e^{i\eta_{2I}}\bigg(e^{\frac{\Delta_{52}+\rho_{12}}{2}}\cosh(\eta_{1R}+\eta_{3R}+\frac{\psi_1}{2})+e^{\frac{\Delta_{12}+\Delta_{22}}{2}}\cosh(\eta_{1R}-\eta_{3R}+\frac{\psi_2}{2})\bigg),\nonumber\\\nonumber
\hspace{-2.5cm}q_3&=&\frac{1}{f}e^{i\eta_{3I}}\bigg(e^{\frac{\Delta_{53}+\rho_{13}}{2}}\cosh(\eta_{1R}+\eta_{2R}+\frac{\chi_1}{2})+e^{\frac{\Delta_{13}+\Delta_{23}}{2}}\cosh(\eta_{1R}-\eta_{2R}+\frac{\chi_2}{2})\bigg),\nonumber\\
\hspace{-2.5cm}f&=&e^{\frac{\delta_{7}}{2}}\cosh(\eta_{1R}+\eta_{2R}+\eta_{3R}+\frac{\delta_{7}}{2})+e^{\frac{\delta_{1}+\delta_{6}}{2}}\cosh(\eta_{1R}-\eta_{2R}-\eta_{3R}+\frac{\delta_{1}-\delta_{6}}{2})\nonumber\\
\hspace{-2.5cm}&&+e^{\frac{\delta_{2}+\delta_{5}}{2}}\cosh(\eta_{2R}-\eta_{1R}-\eta_{3R}+\frac{\delta_{2}-\delta_{5}}{2})+e^{\frac{\delta_{3}+\delta_{4}}{2}}\cosh(\eta_{3R}-\eta_{1R}-\eta_{2R}+\frac{\delta_{3}-\delta_{4}}{2}),~~~\label{tri}
\end{eqnarray}    
where $\eta_{jR}=k_{jR}(t-2k_{jI}z)$, $j=1,2,3$, $\phi_1=\Delta_{51}-\rho_{11}$, $\phi_2=\Delta_{11}-\Delta_{21}$, $\psi_1=\Delta_{52}-\rho_{12}$, $\psi_2=\Delta_{12}-\Delta_{22}$, $\chi_1=\Delta_{53}-\rho_{13}$, $\chi_2=\Delta_{13}-\Delta_{23}$, $\rho_{1j}=\alpha_{1}^{(j)}$, $j=1,2,3$, and the other constants given above are $e^{\delta_{1}}=\frac{|\alpha_{1}^{(1)}|^2}{\Lambda_{11}}$, $e^{\delta_{2}}=\frac{|\alpha_{1}^{(2)}|^2}{\Lambda_{22}}$, $e^{\delta_{3}}=\frac{|\alpha_{1}^{(3)}|^2}{\Lambda_{33}}$, $e^{\Delta_{11}}=\frac{\alpha_{1}^{(1)}\varrho_{12}}{\lambda_{12}}e^{\delta_{2}}$, $e^{\Delta_{21}}=\frac{\alpha_{1}^{(1)}\varrho_{13}}{\lambda_{13}}e^{\delta_{3}}$, $e^{\Delta_{12}}=-\frac{\alpha_{1}^{(2)}\varrho_{13}}{\lambda_{12}^*}e^{\delta_{1}}$, $e^{\Delta_{22}}=\frac{\alpha_{1}^{(2)}\varrho_{23}}{\lambda_{23}}e^{\delta_{3}}$, $e^{\Delta_{13}}=-\frac{\alpha_{1}^{(3)}\varrho_{13}}{\lambda_{13}^*}e^{\delta_{1}}$, $e^{\Delta_{23}}=-\frac{\alpha_{1}^{(3)}\varrho_{23}}{\lambda_{23}^*}e^{\delta_{2}}$, $e^{\delta_{4}}=\frac{|\varrho_{12}|^2}{|\lambda_{12}|^2}e^{\delta_{1}+\delta_{2}}$, $e^{\delta_{5}}=\frac{|\varrho_{13}|^2}{|\lambda_{13}|^2}e^{\delta_{1}+\delta_{3}}$, $e^{\delta_{6}}=\frac{|\varrho_{23}|^2}{|\lambda_{23}|^2}e^{\delta_{2}+\delta_{3}}$, $e^{\delta_{7}}=\frac{|\varrho_{12}|^2|\varrho_{13}|^2|\varrho_{23}|^2}{|\lambda_{12}|^2|\lambda_{13}|^2|\lambda_{23}|^2}e^{\delta_{1}+\delta_{2}+\delta_{3}}$, $e^{\Delta_{51}}=\frac{\alpha_{1}^{(1)}\varrho_{12}\varrho_{13}|\varrho_{23}|^2}{\lambda_{12}\lambda_{13}|\lambda_{23}|^2}e^{\delta_{2}+\delta_{3}}$, $e^{\Delta_{52}}=-\frac{\alpha_{1}^{(2)}\varrho_{12}|\varrho_{13}|^2\varrho_{23}}{\lambda_{12}^*|\lambda_{13}|^2\lambda_{23}}e^{\delta_{1}+\delta_{3}}$,  $e^{\Delta_{53}}=\frac{\alpha_{1}^{(3)}|\varrho_{12}|^2\varrho_{13}\varrho_{23}}{|\lambda_{12}|^2\lambda_{13}^*\lambda_{23}^*}e^{\delta_{1}+\delta_{2}}$, $\Lambda_{11}=(k_1+k_1^*)^2$, $\Lambda_{22}=(k_2+k_2^*)^2$, $\Lambda_{33}=(k_3+k_3^*)^2$, $\varrho_{12}=(k_1-k_2)$, $\varrho_{13}=(k_1-k_3)$, $\varrho_{23}=(k_2-k_3)$, $\lambda_{12}=(k_1+k_2^*)$, $\lambda_{13}=(k_1+k_3^*)$ and $\lambda_{23}=(k_2+k_3^*)$. The above nontrivial soliton solution is described by six arbitrary complex parameters, $\alpha_{1}^{(j)}$,  $k_{j}$, $j=1,2,3$. As a specific example, we can easily check that such multi-parameter solution admits a novel asymmetric triple-hump profile when we fix the velocity as $k_{1I}=k_{2I}=k_{3I}=0.5$. The other parameter values are chosen as $k_{1R} = 0.53$, $k_{2R} = 0.5$, $k_{3R} = 0.45$, $\alpha_{1}^{(1)} = 0.65 + 0.65i$, $\alpha_{1}^{(2)} = 0.45 - 0.45i$ and $\alpha_{1}^{(3)} = 0.35 + 0.35i$. In Fig. 1(a), we display the  asymmetric triple-hump profiles in all the components for the above choice of parameter values. It is important to note that for the specific choice of parameter values, the solution (\ref{tri}) also exhibits symmetric triple-hump soliton profile.  The symmetric and asymmetric nature of solution (\ref{tri}) can be identified by calculating the following relative separation distances between the solitons of the modes,
\numparts
	\begin{eqnarray}
	\Delta t_{12}=t_1-t_2=\frac{1}{2}\log \frac{|\alpha_1^{(1)}|^2(k_{3R}-k_{1R})(k_{2R}+k_{3R})k_{2R}^2}{|\alpha_1^{(2)}|^2(k_{2R}-k_{3R})(k_{1R}+k_{3R})k_{1R}^2},\label{5a}\\
	\Delta t_{13}=t_1-t_3=\frac{1}{2}\log \frac{|\alpha_1^{(1)}|^2(k_{1R}-k_{2R})(k_{2R}+k_{3R})k_{3R}^2}{|\alpha_1^{(3)}|^2(k_{2R}-k_{3R})(k_{1R}+k_{2R})k_{1R}^2},\label{5b}\\
	\Delta t_{23}=t_2-t_3=\frac{1}{2}\log \frac{|\alpha_1^{(2)}|^2(k_{2R}-k_{1R})(k_{1R}+k_{3R})k_{3R}^2}{|\alpha_1^{(3)}|^2(k_{1R}-k_{3R})(k_{1R}+k_{2R})k_{2R}^2}.\label{5c}
	\end{eqnarray}
\endnumparts
It is evident from Eqs. (\ref{5a})-(\ref{5c}) the solution (\ref{tri}), with $k_{1I}=k_{2I}=k_{3I}$, always admits asymmetric triple-hump profiles when  $\Delta t_{12}=\Delta t_{13}=\Delta t_{23}\neq 0$.  In contrast to this, almost symmetric (not perfect symmetric) triple-hump profile arises in all the modes when the soliton parameters obey the condition,  $\Delta t_{12}=\Delta t_{13}=\Delta t_{23}\rightarrow 0$.
 The double node (or multi-node) formation occurs when the relative velocities among the solitons of the modes, $q_j$'s $j=1,2,3$, do not tend to zero. Such node formation is demonstrated in Fig. 2 for the unequal velocity case (of the modes) in the present $3$-CNLS system.  We wish to point out here that the triple peak power profiles obeying the above relative separation distance condition, both symmetric and asymmetric, could be useful in the launching of the initial signal in binary coding scheme. In the practical situation the initial profiles can vary their shape due to fiber's loss and nonlinear higher order effects. This situation cannot be avoided in a fiber. However, the solution (\ref{tri}) retains the fundamental property, namely the triple-hump soliton profile, of the nondegenerate  soliton during the evolution along the fiber.        
\begin{figure}
\begin{center}
\includegraphics[width=5.0cm]{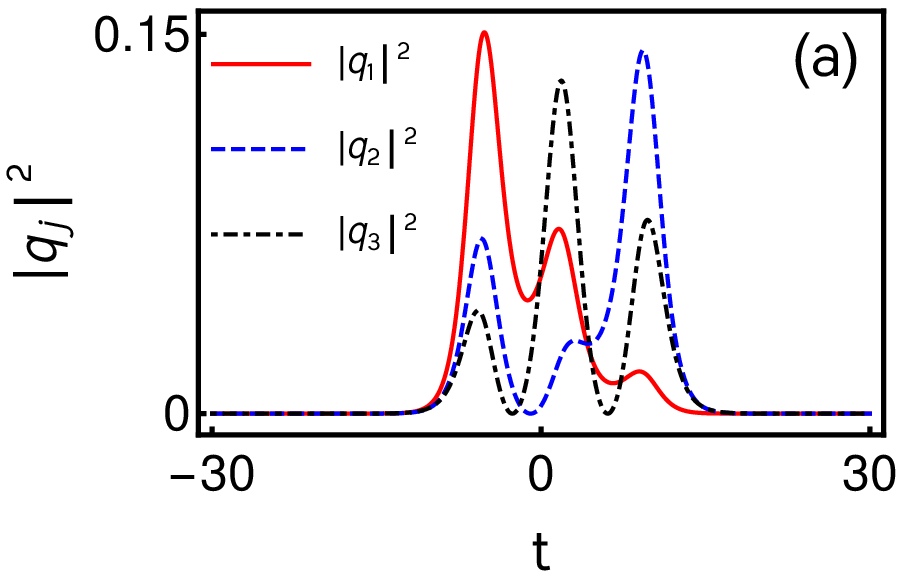}~\includegraphics[width=5.0cm]{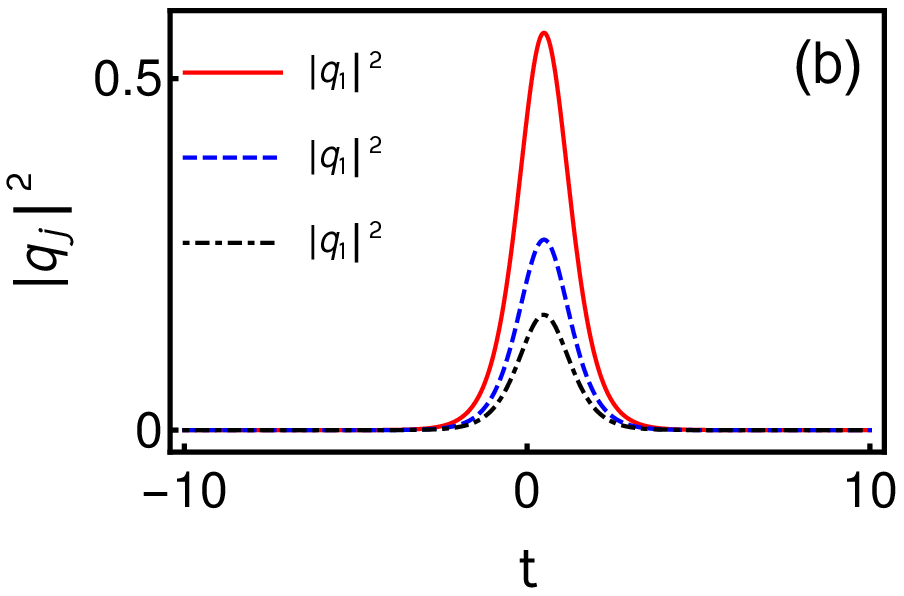}~~\includegraphics[width=5.0cm]{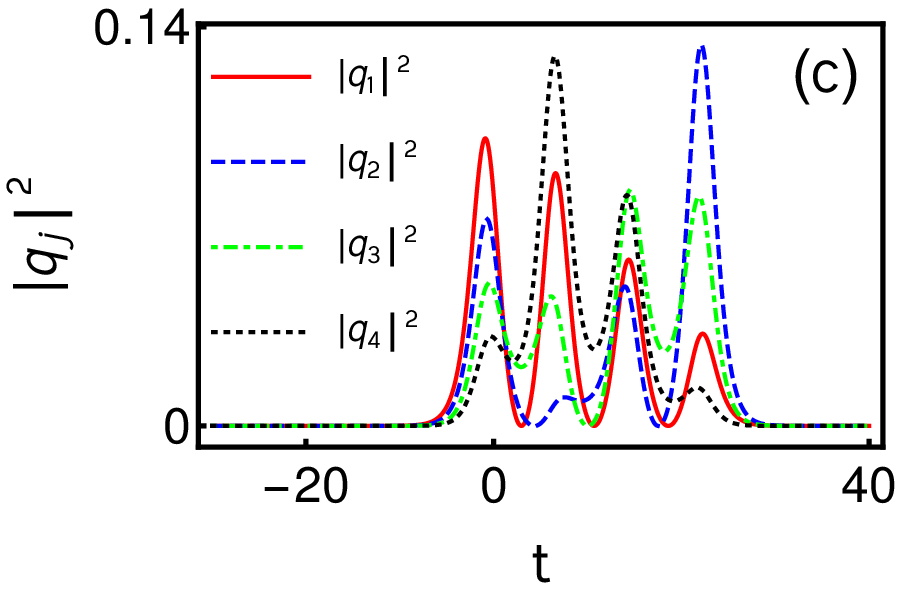}
\caption{(a) denotes triple-hump profiles of nondegenerate fundamental soliton in the $3$-CNLS system and (b) is its corresponding single-humped degenerate soliton profile. (c) represents a quadruple-humped nondegenerate soliton profiles in $4$-CNLS system. The specific values of the soliton parameters are given in the text.}
\label{fig1}
\end{center}
\end{figure}
It is interesting to note that when we impose the condition $k_{1}=k_{2}=k_{3}$ in the solution (\ref{tri}), it turns out to be a single-humped degenerate fundamental soliton for the $3$-CNLS system. This can be seen from Fig. 1(b) for the values $k_1 =k_2 = k_3 = 1+i$, $\alpha_{1}^{(1)} = 0.65 + 0.65i$, $\alpha_{1}^{(2)} = 0.45 - 0.45i$ and $\alpha_{1}^{(3)} = 0.35 + 0.35i$. We note that the $3$-partially coherent soliton or multi-soliton complexes arise from the nondegenerate fundamental soliton solution (\ref{tri}) of the $3$-CNLS system when the soliton parameters are fixed as $\alpha_{1}^{(1)}=e^{\eta_{10}}$, $\alpha_{1}^{(2)}=-e^{\eta_{20}}$, $\alpha_{1}^{(3)}=e^{\eta_{30}}$, $k_{1}=k_{1R}$, $k_{2}=k_{2R}$, $k_{3}=k_{3R}$ and $k_{jI}=0$, $j=1,2,3$, where $\eta_{j0}$, $j=1,2,3$, are considered as real constants \cite{na1, Kanna3}.


 Next we illustrate the multi-hump nature of nondegenerate soliton in the $4$-CNLS system. To obtain such solution one has to proceed with the analysis for the $N=4$ case, as we have described in the above $3$-component case. For brevity, we do not give the details of the final solution due to its complex nature. However, one can easily deduce the form of the solution from the soliton solution of the $N$-component case, Eq. (\ref{eq2}),  as given above.   The final solution contains eight arbitrary complex parameters, namely $\alpha_{1}^{(j)}$ and $k_{j}$, $j=1,2,3,4$. These parameters play a significant role in determining the profile nature of the underlying soliton in the $4$-component case. In general, the nondegenerate one-soliton  solution in the $4$-CNLS system exhibits asymmetric quadruple-hump profile in all the modes. Such novel quadruple-hump profile is displayed in Fig. 1(c) for the parameter values $k_1 = 0.48 + 0.5i$, $k_2 = 0.5 + 0.5i$, $k_3 = 0.53 + 0.5i$, $k_4 = 0.55 + 0.5i$, $\alpha_{1}^{(1)} = 0.65 + 0.65i$, $\alpha_{1}^{(2)} = 0.55 - 0.55i$, $\alpha_{1}^{(3)} = 0.45 + 0.45i$ and $\alpha_{1}^{(4)} = 0.35 - 0.35i$.
\begin{figure}
\begin{center}
\includegraphics[width=15.0cm]{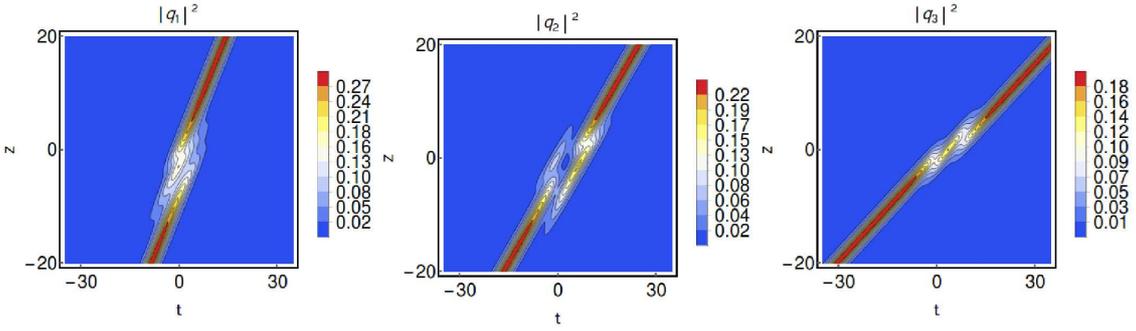}
\caption{Double-node formation in the unequal velocities case in the profile of nondegenerate fundamental soliton in $3$-CNLS system. The parameter values are $k_1 = 0.55 + 0.35i$, $k_2 = 0.5 + 0.5i$, $k_3 = 0.45 + 0.8i$ $\alpha_{1}^{(1)} = 0.65 + 0.65i$, $\alpha_{1}^{(2)} = 0.45 - 0.45i$ and $\alpha_{1}^{(3)} = 0.35 + 0.35i$.}
\label{fig2}
\end{center}
\end{figure} 
We have verified the asymmetric quadruple-hump profile nature by calculating the relative separation distance, $\Delta t_{12}=\Delta t_{13}=\Delta t_{14}\neq 0$. However we do not present their explicit forms due to size limitation of the letter article. It is evident from Figs. 1(a) and 1(c) that the nondegenerate soliton (in $3$, $4$ and also in the arbitrary $N$ ($>4$) CNLS systems) exhibits multi-hump nature. This multi-peak nature can increase the bit-rate in coding the information. Consequently it can help to uplift the flow of data in fiber. In the present $4$-CNLS system case also multi-node forms when the relative velocities of the solitons among the modes do not tend to zero. One can also recover the already known degenerate soliton solution by fixing the condition $k_{1} = k_{2} = k_{3} = k_{4}$ in the final form of nondegenerate soliton solution of the $4$-CNLS system.  

 In the following, we further report the fact that the $N$-CNLS system can also admit very interesting partially nondegenerate soliton solution when the wavenumbers are restricted suitably.  Such partial nondegenerate soliton solutions also exhibit multi-hump profiles (but less than $N$ in number). For instance, here we demonstrate their existence for the $3$ and $4$-CNLS systems and this procedure can be generalized to the $N$-component case in principle. For the 3-component case, the partially nondegenerate soliton solution can be obtained by imposing the condition, $k_{1} = k_{2}$ (or $k_{1} = k_{3}$ or $k_{2} = k_{3}$), on the wave numbers in the solution (\ref{tri}). This restriction reduces the asymmetric triple-hump profile, as depicted in Fig. 1(a), into the asymmetric double-hump intensity profile as displayed in Fig. 3(a) for the choice of parameters $k_1 = k_2=0.5 + 0.5i$,  $k_3 = 0.45+ 0.5i$, $\alpha_{1}^{(1)} = 0.65 + 0.65i$, $\alpha_{1}^{(2)} = 0.45 - 0.45i$ and $\alpha_{1}^{(3)} = 0.35 + 0.35i$.  The partially NDS double-hump profile is described by the following explicit form of solution, deduced from solution (\ref{tri}), \
\begin{eqnarray}
\hspace{-2.6cm}q_1&=&\frac{1}{f}e^{i\eta_{1I}}e^{\frac{\Delta_{21}+\rho_{11}}{2}}\cosh(\eta_{3R}+\frac{\Delta_{21}-\rho_{11}}{2}),~q_3=\frac{1}{f}e^{i\eta_{3I}}e^{\frac{\Delta+\rho_{13}}{2}}\cosh(\eta_{1R}+\frac{\Delta-\rho_{13}}{2}),\nonumber\\
\hspace{-2.6cm}q_2&=&\frac{1}{f}e^{i\eta_{1I}}\bigg(\frac{1}{2}[\cosh(2\eta_{1R}-\eta_{3R}+\Delta_{12})+\sinh(2\eta_{1R}-\eta_{3R}+\Delta_{12})]\nonumber\\
\hspace{-2.6cm}&&+e^{\frac{\Delta_{22}+\rho_{12}}{2}}\cosh(\eta_{3R}+\frac{\Delta_{22}-\rho_{12}}{2})\bigg),\nonumber\\
\hspace{-2.6cm}f&=&e^{\frac{\bar{\delta}_{1}}{2}}\cosh(\eta_{1R}+\eta_{3R}+\frac{\bar{\delta}_{1}}{2})+e^{\frac{\bar{\delta}_{2}+\delta_{3}}{2}}\cosh(\eta_{1R}-\eta_{3R}+\frac{\bar{\delta}_{2}-\delta_{3}}{2}).
\end{eqnarray}   
In the above $e^{\bar{\delta}_{1}}=e^{\delta_5}+e^{\delta_6}$, $e^{\bar{\delta}_{2}}=e^{\delta_1}+e^{\delta_2}$, $e^{\Delta}=e^{\Delta_{13}}+e^{\Delta_{23}}$, $\eta_1=\eta_2=k_1t+ik_1^2z$,  $\eta_3=k_3t+ik_3^2z$ and the other constants are deduced from the constants of the solution (\ref{tri}) by imposing the condition $k_1=k_2$ in them. We point out that one can get the degenerate soliton solution by imposing the restriction further on the wavenumbers, that is as we mentioned above $k_1=k_2=k_3$ leads to completely degenerate soliton solution.      
\begin{figure}
\begin{center}
\includegraphics[width=5.0cm]{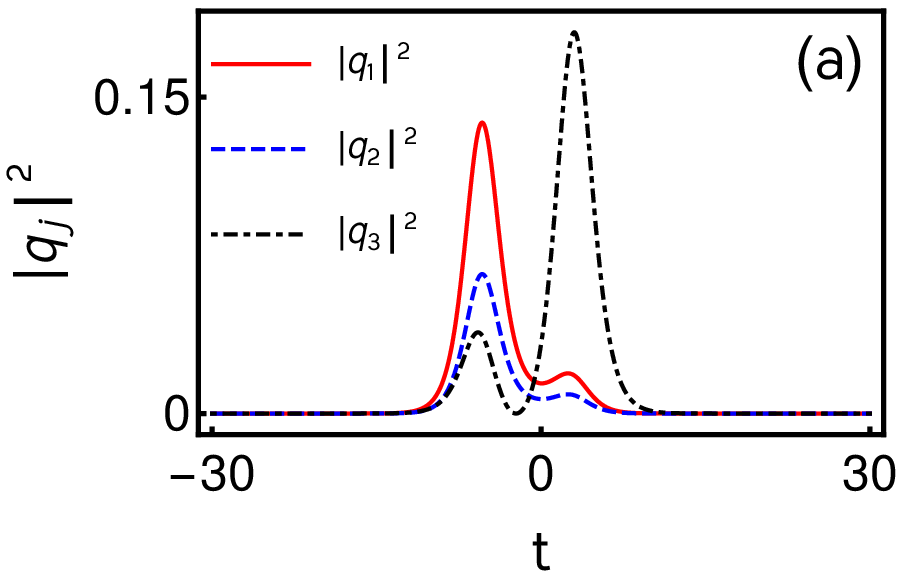}~\includegraphics[width=5.0cm]{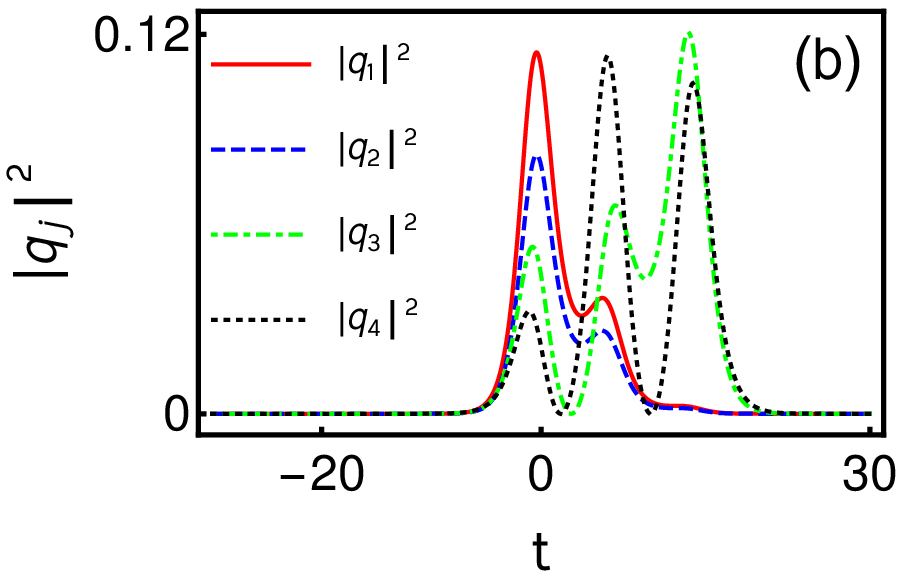}~\includegraphics[width=5.0cm]{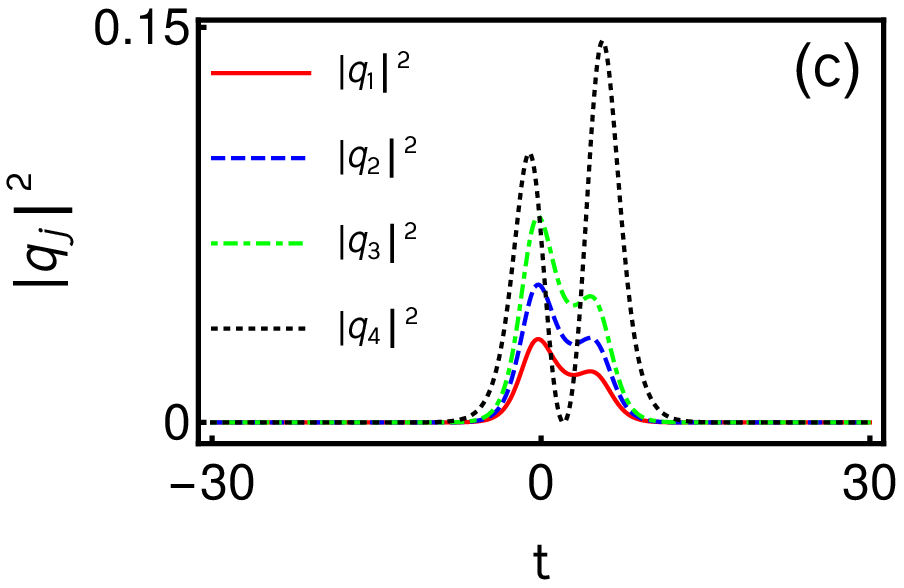}
\caption{(a) denotes double-humped profile of the partially nondegenerate one soliton solution of $3$-CNLS system. (b) and (c) represent triple and double-humped profiles of partially nondegenerate soliton solution of $4$-CNLS system when the conditions  $k_1=k_2$ and $k_1=k_2=k_3$ on wavenumbers are imposed, respectively.  }
\label{fig3}
\end{center}
\end{figure}
It is important to note that partially nondegenerate soliton solution of the 3-CNLSE can exhibit only upto double hump profile in all the three modes due to the degeneracy among the modes and the nature of this solution is controlled by five arbitrary complex parameters.     

Similarly, for the 4-CNLS equation, partially nondegenerate soliton solution can be deduced from  the solution (\ref{eq2}) of $N$-component case. However, due to the complex nature of the resultant solution we do not present the expression here. Very interestingly such solution provides the following three possibilities: (i). $k_1=k_2$, (ii). $k_1=k_2=k_3$ and (iii) $k_1=k_2=k_3=k_4$. The quadruple-hump soliton profile of the $4$-CNLS system becomes a triple-hump profile when we consider the first possibility, $k_1=k_2$. This triple-humped partially nondegenerate soliton solution is diplayed in Fig. 3(b) for $k_1 = k_2=0.55 + 0.5i $, $k_3 = 0.5 + 0.5i $, $k_4 = 0.45 + 0.5i $, $\alpha_{1}^{(1)} = 0.65 + 0.65i $, $\alpha_{1}^{(2)} = 0.55 - 0.55i $, $\alpha_{1}^{(3)} = 0.45 + 0.45i $ and $\alpha_{1}^{(4)} = 0.35 - 0.35i $. In contrast to the latter, we observe that the double-hump soliton profile emerges while considering the second possibility, $k_1=k_2=k_3$, in the full nondegenerate form of solution of the $4$-CNLS system. Such double-humped partially NDS solution profile is depicted in Fig. 3(c) for the values $k_1 = k_2=k_3=0.55 + 0.5i $, $k_4 = 0.45 + 0.5i $, $\alpha_{1}^{(1)} = 0.35 + 0.35i $, $\alpha_{1}^{(2)} = 0.45 + 0.45i $, $\alpha_{1}^{(3)} = 0.55 + 0.55i $ and $\alpha_{1}^{(4)} = 0.65 - 0.65i $. The final possibilty, $k_1=k_2=k_3=k_4$, corresponds to complete degeneracy. This choice brings out the completely degenerate soliton solution for the $4$-CNLS system. In general, for the $N$-component case, one would expect $N-1$ possibilities of choices of wave numbers. Out of these choices a single-humped complete degenerate soliton solution (\ref{deg}) arises if all the wavenumbers are equal, $k_1=k_2=...=k_n$, whereas the partial nondegeneracy appears from out of the remaining $N-2$  possibilities. Such partial nondegeneracy would bring out multi-hump profiles as we have illustrated above for the $3$ and $4$ component cases.             
 
\begin{figure}
	\begin{center}
		\includegraphics[width=15cm]{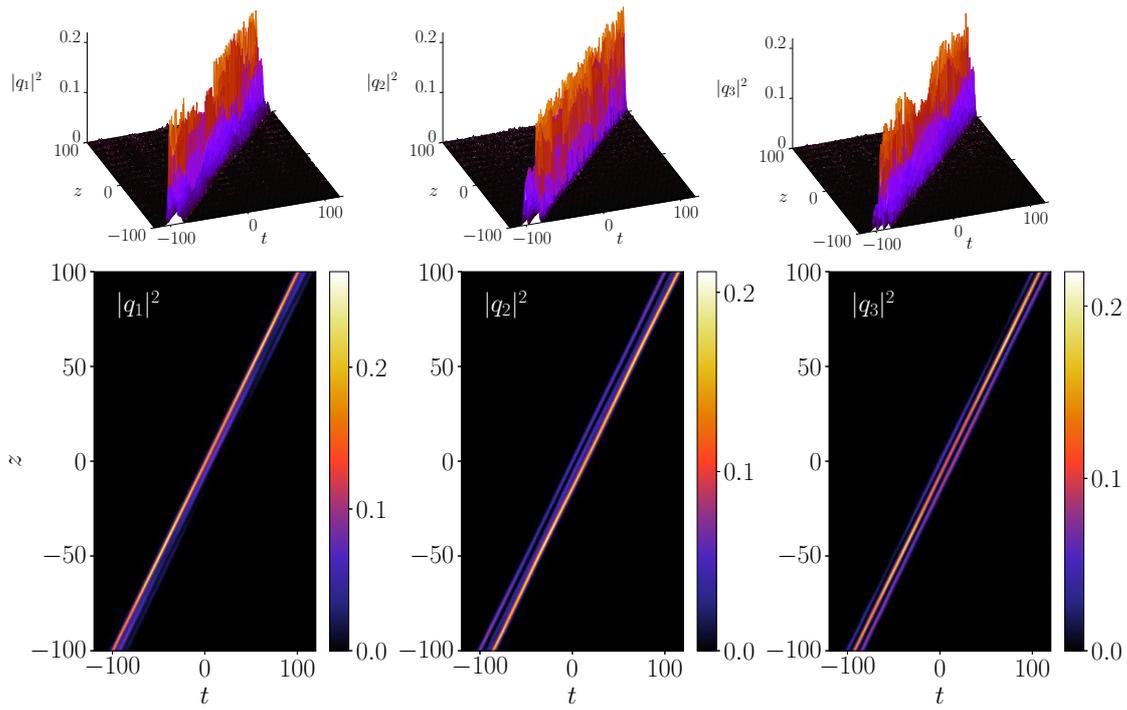}
		\caption{ Numerical plots for the asymmetric nondegenerate triple hump soliton profile with $5\%$ of white noise as perturbation. Top panel denotes the triple-hump profile of 3-dimensional surface plot and the bottom panel represents the corresponding density plots.  The soliton parameters correspond to Fig. 1(a).}
		\label{fig4}
	\end{center}
\end{figure} 
\begin{figure}
	\begin{center}
		\includegraphics[width=15cm]{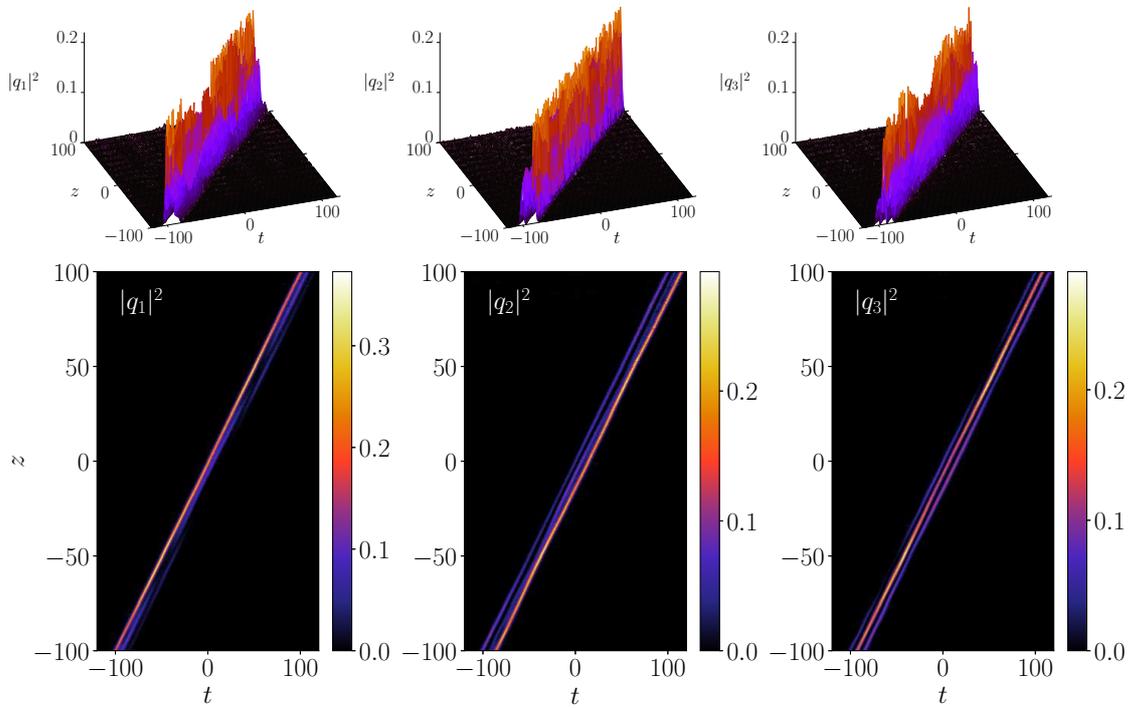}
		\caption{Numerical plots for the asymmetric nondegenerate triple hump soliton profile with $10\%$ of white noise as perturbation. }
		\label{fig5}
	\end{center}
\end{figure} 
We also wish to point out the stability nature of the triple-humped nondegenerate fundamental soliton solution (\ref{tri}) of the $3$-CNLS system as an example. In order to do this, we consider the Crank-Nicolson numerical algorithm \cite{pm} with different percentages of white noise as perturbations to the initial profiles. The initial profiles are considered in the numerical analysis as $q_{j}(-100,t)=[1+A \zeta(t)]q_{j,-100}(t)$, $j=1,2,3$, where $q_{j,-100}(t)$, $j=1,2,3$, are the initial profiles obtained from the solution (\ref{tri}) at $z=-100$. Here, $A$ is the amplitude of the white noise which is generated from the random numbers in the interval $[-1,1]$ and $\zeta(t)$ is the noise function. The space and time step sizes are fixed in the  numerical calculation, respectively, as  $dz=0.1$ and $dt=0.2$. We also fix the domain range values for both $t$ and $z$ as $ [-100,100]$. The triple-hump profile nature survives during the evolution even for $5\%$ and $10\%$ of white noise perturbations except for minor changes in the amplitude part. This is clearly demonstrated in figures 4 and 5. These figures ensure the stability of triple-humped nondegenerate soliton against perturbations of white noise. One can extend this analysis for even longer ranges of time and space without much difficulty. Similarly, we have also confirmed the stability of asymmetric quadruple-hump nondegenerate soliton of the $4$-CNLS system as well.

 In this paper, we reported the existence of nondegenerate fundamental soliton solution for the $N$-coupled nonlinear Schr\"{o}dinger equations (1). This new class of solitons exhibit multi-hump nature among all the modes. The existence of such special multi-humped profiles is demonstrated explicitly by considering the nondegenerate soliton solution for the $3$ and $4$ component cases. Very interestingly we have also shown the existence of partially nondegenerate soliton solutions by restricting the wave numbers suitably. The already known energy exchanging degenerate class of vector bright solitons is shown as a sub-case by imposing specific restriction on the wave numbers. Finally, the stability of multi-humped nondegenerate fundamental soliton has also been verified numerically. In a subequent work we have planned to report the interesting collision properties of these nondegenerate solitons. We believe that the existence of multi-peak power nature in the nondegenerate fundamental soliton in multi-mode optical fibers may be relevant to increase the data stream in multi-level optical communication applications.

The authors are  thankful to Prof. P. Muruganandam, Department of Physics, Bharathidasan University, Tiruchirapalli - 620 024 for verifying the multi-hump nature of nondegenerate solitons numerically with white noise as perturbation. R. R., S. S. and M. L. acknowledge the financial support in the form of DST-SERB Distinguished Fellowship program to M. L. under the Grant No. SB/DF/04/2017.  \\

\end{document}